\documentclass[%
 aip,
 amsmath,amssymb,
 reprint,%
]{revtex4-1}

\usepackage{graphicx}
\usepackage{dcolumn}
\usepackage{bm}
\usepackage{caption}
\usepackage{subcaption}
\usepackage[utf8]{inputenc}
\usepackage[T1]{fontenc}
\usepackage{mathptmx}
\usepackage{etoolbox}
\usepackage[export]{adjustbox}
\makeatletter
\def\@email#1#2{%
 \endgroup
 \patchcmd{\titleblock@produce}
  {\frontmatter@RRAPformat}
  {\frontmatter@RRAPformat{\produce@RRAP{*#1\href{mailto:#2}{#2}}}\frontmatter@RRAPformat}
  {}{}
}%
\makeatother
\begin{document}

\preprint{AIP/123-QED}

\title{Analysis of two-terminal perovskite/silicon tandem solar cells with differing texture structure, perovskite carrier lifetime and tunneling junction quality}

\author{Chun-Hao Hsieh}
\affiliation{Graduate Institute of Photonics and Optoelectronics and Department of Electrical Engineering,\\ National Taiwan University, Taipei 10617, Taiwan}
\author{Jun-Yu Huang}
\affiliation{Graduate Institute of Photonics and Optoelectronics and Department of Electrical Engineering,\\ National Taiwan University, Taipei 10617, Taiwan}
\author{Yuh-Renn Wu}
\email{yrwu@ntu.edu.tw}
\affiliation{Graduate Institute of Photonics and Optoelectronics and Department of Electrical Engineering,\\ National Taiwan University, Taipei 10617, Taiwan}

%

\date{\today}

\begin{abstract}
Presented here is the optimization of a planar two-terminal perovskite/silicon tandem solar cell with a texture structure. The developed simulation model is fitted to published experimental results, and the importance of current matching in the two-terminal structure is discussed. With the texture structure optimized and considering current matching, the optimal texture structure improves $J_{sc}$ from 17.9~mA/cm$^2$ to 20.87~mA/cm$^2$ compared to the planar structure, as well as improving the power conversion efficiency from 25.8\% to 35.9\%. Furthermore, if the quality of the perovskite thin film and tunneling junction efficiency with a smaller voltage penalty can be improved, then the efficiency can be further improved to 38.13\%. This indicates that this tandem solar cell still has much room for improvement.
\end{abstract}

\maketitle

\section{Introduction}

The power conversion efficiency (PCE) of single-junction perovskite (PVSK) solar cells has now surpassed 20\%,\cite{PSC1,PSC2,PSC3,PSC4,PSC5,PSC6,Ref13,Ref14} thereby offering an excellent opportunity for further development of tandem solar cells (TSCs). In comparison with multi-junction TSCs, stacking more layers will increase the manufacturing costs and the loss of the tunneling junction between each absorption layer, so a stack of two layers of materials is currently considered a more feasible structure. Perovskite/Si TSCs are believed to be the key low-cost technology for breaking through the single-junction efficiency limits. Some teams have proposed perovskite/semiconductor-based TSCs,\cite{Ref15,Ref16,Ref17,Ref18,Ref19,Ref20,Ref21,Ref22} and the US National Renewal Energy Laboratory has confirmed that their PCE could exceed 30\%.\cite{KAUST-32.5,HZB-32.5,EPFL-31.25,planar-Voc1.9} However, the device design for TSCs is more difficult than that for conventional single-layer solar cells for the following two main reasons. 1) A TSC is physically equivalent to two single-material solar cells connected in series, so current matching must be achieved when designing the absorption spectrum; excess current causes losses, so the distribution of light absorption is an important issue. The structure of perovskite materials can be adjusted via the solution composition,\cite{Ref23,PVKcomposotion,PVKcomposotion2} so it must be optimized for different perovskite materials and the thickness adjusted. In addition, surface structures can also optimize light absorption at specific wavelengths\cite{texturesurface1,texturesurface2,texturesurface3}. And if the textured structure operates on a smaller scale than the incident wavelength, the textured interface functions as an effective medium characterized by a smoothly varying refractive index. With this continuous gradient in refractive index, incident light undergoes no abrupt transitions, leading to reduced reflectance and consequently higher transmittance.\cite{kim2020surface}  2) Between the top and bottom cells, a tunneling junction must be designed to allow electrons and holes to pass through easily. For tunneling junctions, high-concentration doping is mainly used to form the interfaces, which can lead to tunneling by electrons and holes.\cite{Ref24,TJ,TJ2} For the device design, the light absorption of the top and bottom cells must be considered separately for current matching, so numerical simulations must be performed before component fabrication.

A two-terminal (2T) solar cell has a relatively complicated design, and we must consider the current matching of each layer. This includes the optical-absorption current matching and the need to consider the current-matching conditions after considering the nonradiative loss of the two absorption layers. Herein, we optimize these structures and assess the possibility of optimizing this TSC. This includes (i) optical optimization for the current-matching conditions using a rigorous coupled-wave analysis (RCWA) solver and (ii) electrical optimization (including the nonradiative loss and tunneling junction issues) using a 2D finite-element Poisson and drift-diffusion (Poisson-DD) solver.

\section{Methodology}\label{methods}

To simulate effectively the performance of perovskite/Si TSCs, we use two main simulation models. For the optical modeling, we use the RCWA method to solve Maxwell's equations, and for modeling the electrical properties, we use the 2D Poisson-DD solver developed in our laboratory\cite{huang2022influences,chen2017design,10.1063/5.0088593}.

\subsection{Simulation of optical properties}\label{Simulation of Optical Properties}

The 2D RCWA method is used to simulate the absorption of sunlight in the device; this includes TE and TM modes, and in this study the light source is the AM1.5G solar spectrum. The RCWA algorithm solves Maxwell's equations [see \eqref{eq1} and \eqref{eq2}] in the frequency domain, and this method is used widely for optical simulation on periodic structures. The magnetic field is obtained as follows: 
\begin{equation}
    H_z(x,z)=\frac{-1}{i\omega\mu}\frac{\partial E_y}{\partial x},
    \label{eq1}
\end{equation}
\begin{equation}
    H_x(x,z)=\frac{1}{i\omega\mu}\frac{\partial E_y}{\partial z}.
    \label{eq2}
\end{equation}

For TM polarization, Maxwell's equations are used to obtain the electric field as follows:
\begin{equation}
    E_z(x,z)=\frac{-1}{i\omega\mu\epsilon_r}\frac{\partial H_y}{\partial x},
    \label{eq3}
\end{equation}
\begin{equation}
    E_x(x,z)=\frac{1}{i\omega\mu\epsilon_r}\frac{\partial H_y}{\partial z},
    \label{eq4}
\end{equation}
where $\omega$ is the angular frequency, $\mu$ is the permeability, and $\epsilon_r$ is the relative permittivity. Simulating the propagation of waves accurately requires a dense mesh for the triangular texture structure, thereby requiring significant memory resources if the entire TSC device is to be modeled. To mitigate this computational burden, periodic boundary conditions are employed.

\begin{figure}[h]
\includegraphics[width=0.49\textwidth]{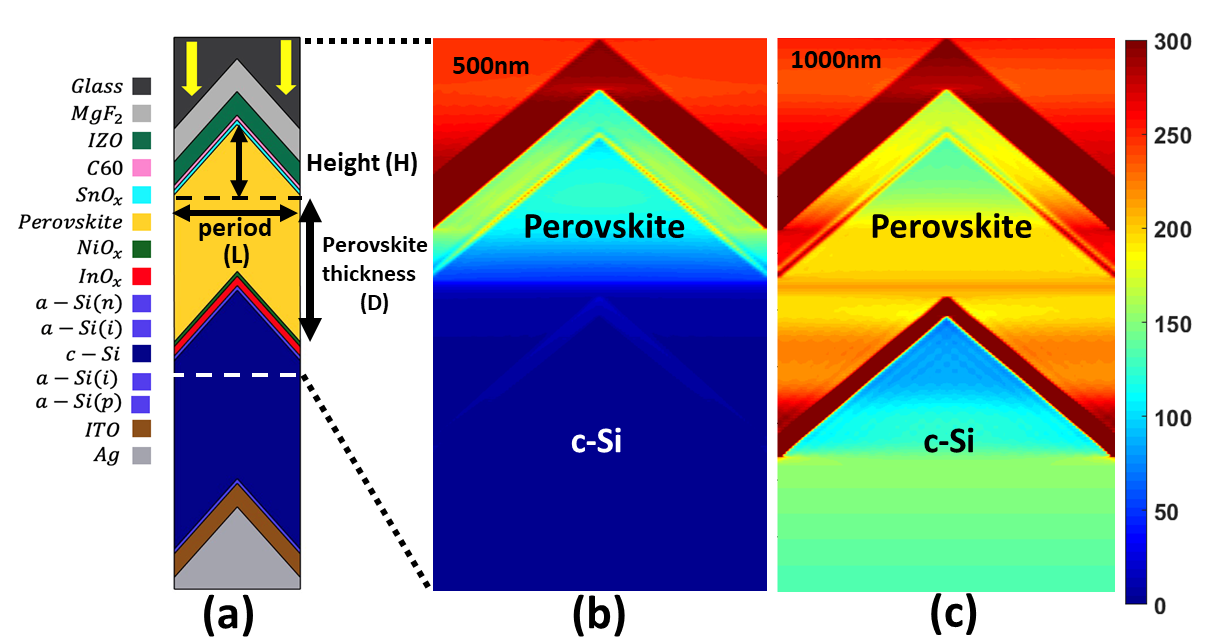}
\caption{(a) Textured perovskite/c-Si tandem solar cell (TSC) structure used in simulations. Distributions of optical field $|E|$ for $\lambda=$ (b) 500~nm and (c) 1000~nm.}
\label{fig:1}
\end{figure}

\begin{figure}[h]
\includegraphics[width=0.49\textwidth]{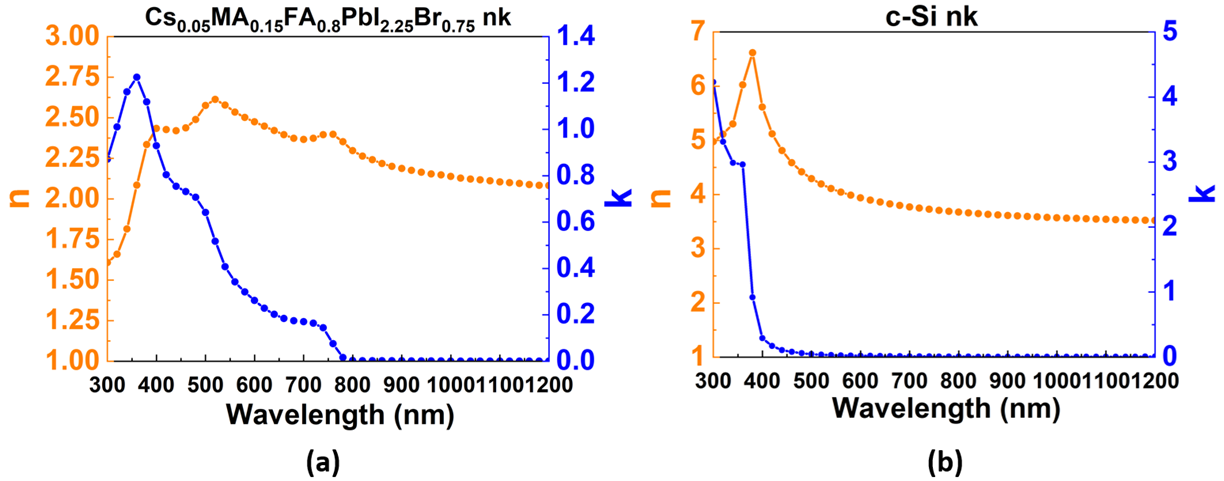}
\caption{Refractive index and extinction coefficient of (a) perovskite and (b) c-Si.}
\label{fig:2}
\end{figure}

Furthermore, the simulated structure is a fully textured one, as shown in Fig.~\ref{fig:1}(a). To find the best PCE, the structure of the device is optimized by trying different values for the period $L$ and height $H$. In addition, the materials selected for our simulated structure are based on experimental results for perovskite/c-Si TSCs published by Hou \textit{et~al.}\cite{Ref25} in 2020. Each layer of material from top to bottom of the device is 
$\text{MgF}_{\text{2}}$ (140~nm) for the antireflection film, IZO (100~nm) for the top electrode, $\text{C}_{\text{60}}$ (20~nm), $\text{SnO}_{\text{x}}$ (20~nm), for the electron transport layer (ETL), perovskite ($\text{Cs}_{\text{0.05}}$$\text{MA}_{\text{0.15}}$$\text{FA}_{\text{0.8}}$$\text{PbI}_{\text{2.25}}$$\text{Br}_{\text{0.75}}$) for the top absorption layer, $\text{NiO}_{\text{x}}$ (17~nm) for the hole transport layer (HTL), $\text{InO}_{\text{x}}$ (40~nm) for the recombination junction, n-type a-Si (8~nm) and intrinsic a-Si (13~nm) for the ETL, crystalline silicon (c-Si; 250~$\mu$m) for the bottom absorption layer, intrinsic a-Si (13~nm) and p-type a-Si (6~nm) for the HTL, ITO (150~nm) and Ag (250~nm) for the back electrodes. Figures~\ref{fig:2}(a) and \ref{fig:2}(b) show the refractive index and extinction coefficient of the two main absorption layers used in the RCWA simulation.

The optical field distributions in Figs.~\ref{fig:1}(b) and \ref{fig:1}(c) reveal that short-wavelength (500~nm) light is absorbed after passing through the perovskite layer and is not transmitted to the bottom cell, whereas long-wavelength (1000~nm) light passes through the perovskite layer and continues to be transmitted to the silicon layer for absorption.

\subsection{Simulation of electrical properties}\label{Simulation of Electrical Properties}

The 2D Poisson-DD solver\cite{huang2022influences,huang2019optimization,chen2017design} is used to analyze the characteristics of perovskite/Si TSCs. Having obtained the optical field distribution, the TE and TM generation rates of the electron--hole pairs in the overall device are calculated as follows: 
\begin{equation}
    G_{opt,TE}=\frac{1}{\hbar\omega}Re(\nabla\cdot P)=\frac{nk}{\hbar}\epsilon_0|\Vec{E}|^2,
    \label{eq5}
\end{equation}
\begin{equation}
    G_{opt,TM}=\frac{1}{\hbar\omega}Re(\nabla\cdot P)=\frac{nk}{\hbar}\frac{1}{N^2}\mu_0|\Vec{H}|^2,
    \label{eq6}
\end{equation}
\begin{equation}
    G_{opt}=\frac{G_{opt,TE}+G_{opt,TM}}{2},
    \label{eq7}
\end{equation}
where $G_{opt,TE}$ and $G_{opt,TM}$ are the generation rates determined by the electric field and magnetic field separately, $\omega$ is the photon frequency, $P$ is the Poynting vector, $n$ is the refractive index, $k$ is the extinction coefficient, and $\Vec{E}$ and $\Vec{H}$ are the steady-state electric field and magnetic field obtained from the RCWA solver as already mentioned. 

\begin{figure}[h]
  \includegraphics[width=0.49\textwidth]{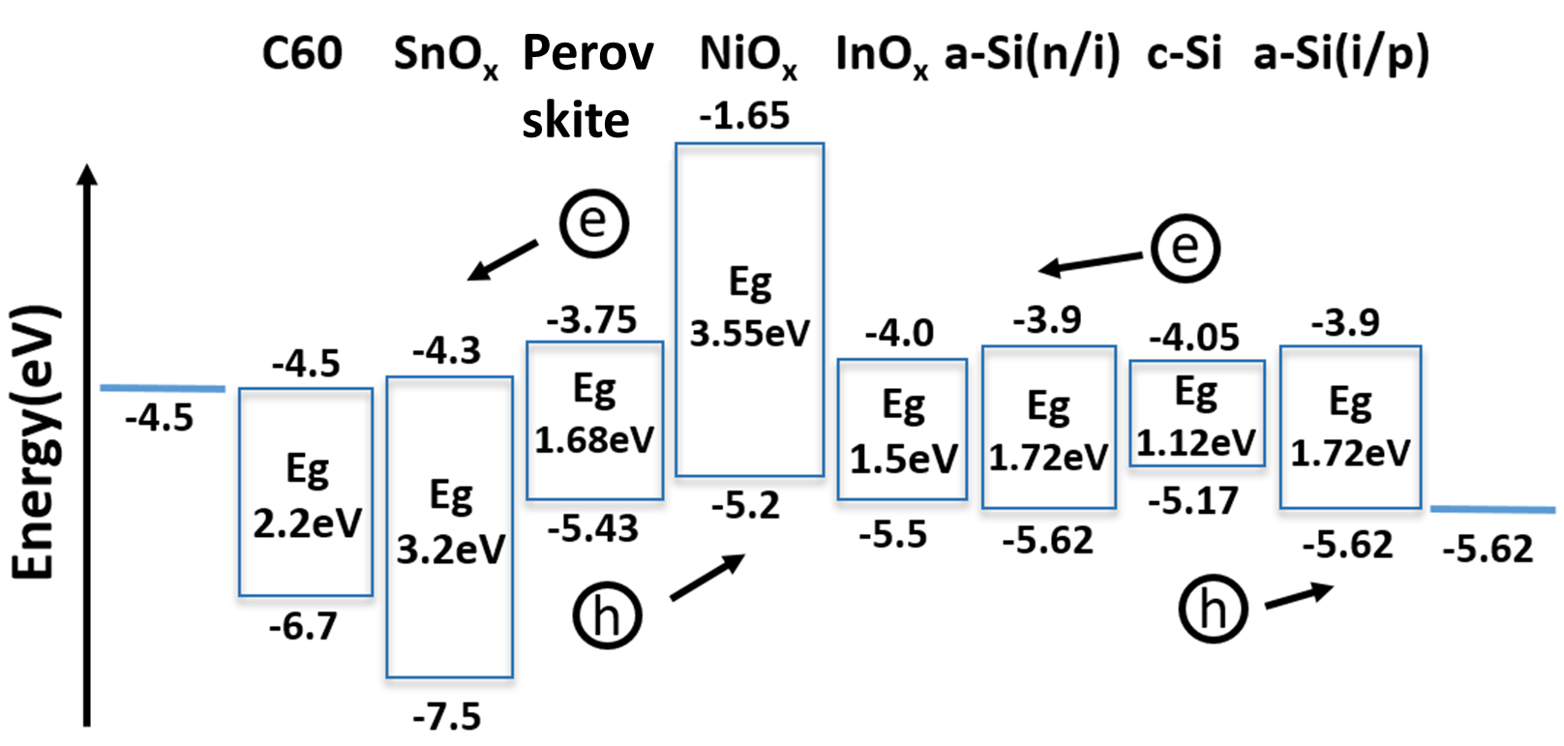}
\caption{Band alignment of simulation materials, including bandgap and electron affinity.}
\label{fig:3}
\end{figure}

\begin{figure}[h]
  \includegraphics[width=0.35\textwidth]{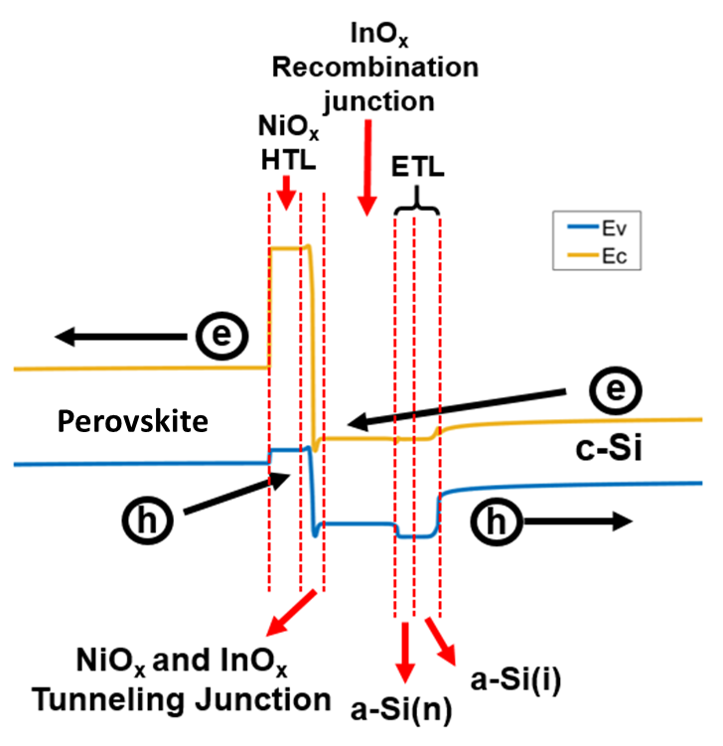}
\caption{Band diagram of tunneling junction.}
\label{fig:4}
\end{figure}

$G_{opt}$ is then passed to the 2D Poisson-DD solver to simulate the electrical characteristics of the device. The main governing equations are as follows:
\begin{equation}
    \nabla_r\cdot(\epsilon\nabla_rV(r))=q(n-p+N_a^--N_d^++\dots),
    \label{eq8}
\end{equation}
\begin{equation}
    J_n=-q\mu_n n(r)\nabla_rV(r)+qD_n\nabla_rn(r),
    \label{eq9}
\end{equation}
\begin{equation}
    J_p=-q\mu_p p(r)\nabla_rV(r)-qD_p\nabla_rp(r),
    \label{eq10}
\end{equation}
\begin{equation}
    \nabla_r(J_{n,p})=(R-G_{opt}),
    \label{eq11}
\end{equation}
where $\epsilon$ is the dielectric constant of the element at different positions, $V$ is the potential of the element, and $n$ and $p$ are the electron and hole densities, respectively. For the drift-diffusion equations, $J_n$ and $J_p$ are the electron and hole currents, respectively, $\mu_n$ and $\mu_p$ are the electron and hole mobilities, respectively, and $D_n$ and $D_p$ are the electron and hole diffusion coefficients, respectively. $R$ is the recombination rate [see \eqref{eq12} and \eqref{eq13}] determined by the Shockley--Read--Hall (SRH) nonradiative recombination and radiative recombination coefficient $B$:
\begin{equation}
    R=SRH+Bnp+C(n^2p+np^2),
    \label{eq12}
\end{equation}
\begin{equation}
    SRH=\frac{np-n_i^2}{\tau_n(p+n_iexp(\frac{E_i-E_t}{k_BT}))+\tau_p(n+n_iexp(\frac{E_t-E_i}{k_BT}))},
    \label{eq13}
\end{equation}
where $\tau_n$ and $\tau_p$ are the nonradiative carrier lifetimes, $C$ is the Auger coefficient, and $n_i$ is the intrinsic carrier concentration. Fig.~\ref{fig:3} shows the bandgap and electron affinity of each material that we input into the 2D Poisson-DD solver for the electrical simulations.

For the carrier exchange between the top and bottom cells, as shown in Fig.~\ref{fig:4}, a tunneling junction is designed between the perovskite and c-Si absorption layers. To model the tunneling junction, the defect state-assisted tunneling recombination mechanism needs to be incorporated into the simulation to make the tunneling junction work. Hence, the tail state models at the tunneling junction are applied to study the tunneling effects. These models assume a Gaussian distribution density of defect states inside the bandgap region where carriers can hop into these defect states in the tunneling depletion region and recombine. This approach is used to simulate the carrier tunneling through the junction. The detailed model related to the tail state model we developed is in Ref. \cite{huang2020analysis, huang2020revealing, huang2023numerical}.

\section{Results and Discussion}

\begin{figure}[h]
  \includegraphics[width=0.45\textwidth,height=0.35\textwidth]{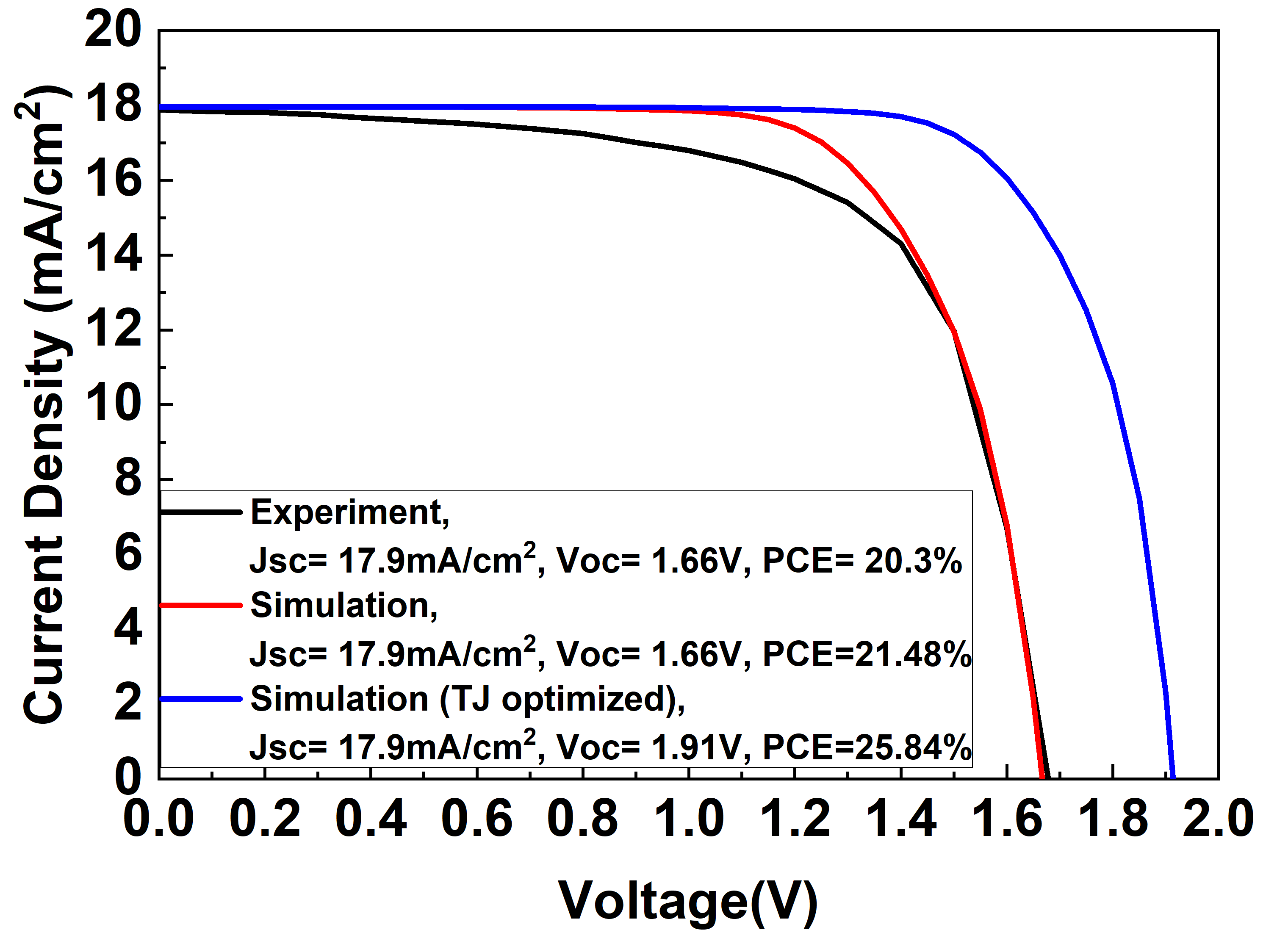}
\caption{Planar-structure $J$--$V$ curves from experiment and simulations.}
\label{fig:5}
\end{figure}

Before optimizing the texture structure of the device, we seek to calibrate our electrical results to the experimental results for the planar structure, and the characteristic parameters are compared in Table~\ref{table:1}. As shown in Fig.~\ref{fig:5}, the fitting of the characteristic parameters to the experimental values is quite close. After the fitting, we realized that the $V_{oc}$ value in the referenced experimental paper is quite small compared to those for other planar perovskite/Si TSCs, in which $V_{oc}$ can reach 1.9~V.\cite{planar-Voc1.9} So, we improved the tunneling junction quality by increasing the defect-assisted tunneling concentration $N_t$ to reach higher $V_{oc}$, as shown in Fig.~\ref{fig:5}.

\begin{figure}[h]
  \includegraphics[width=0.49\textwidth,height=0.36\textwidth]{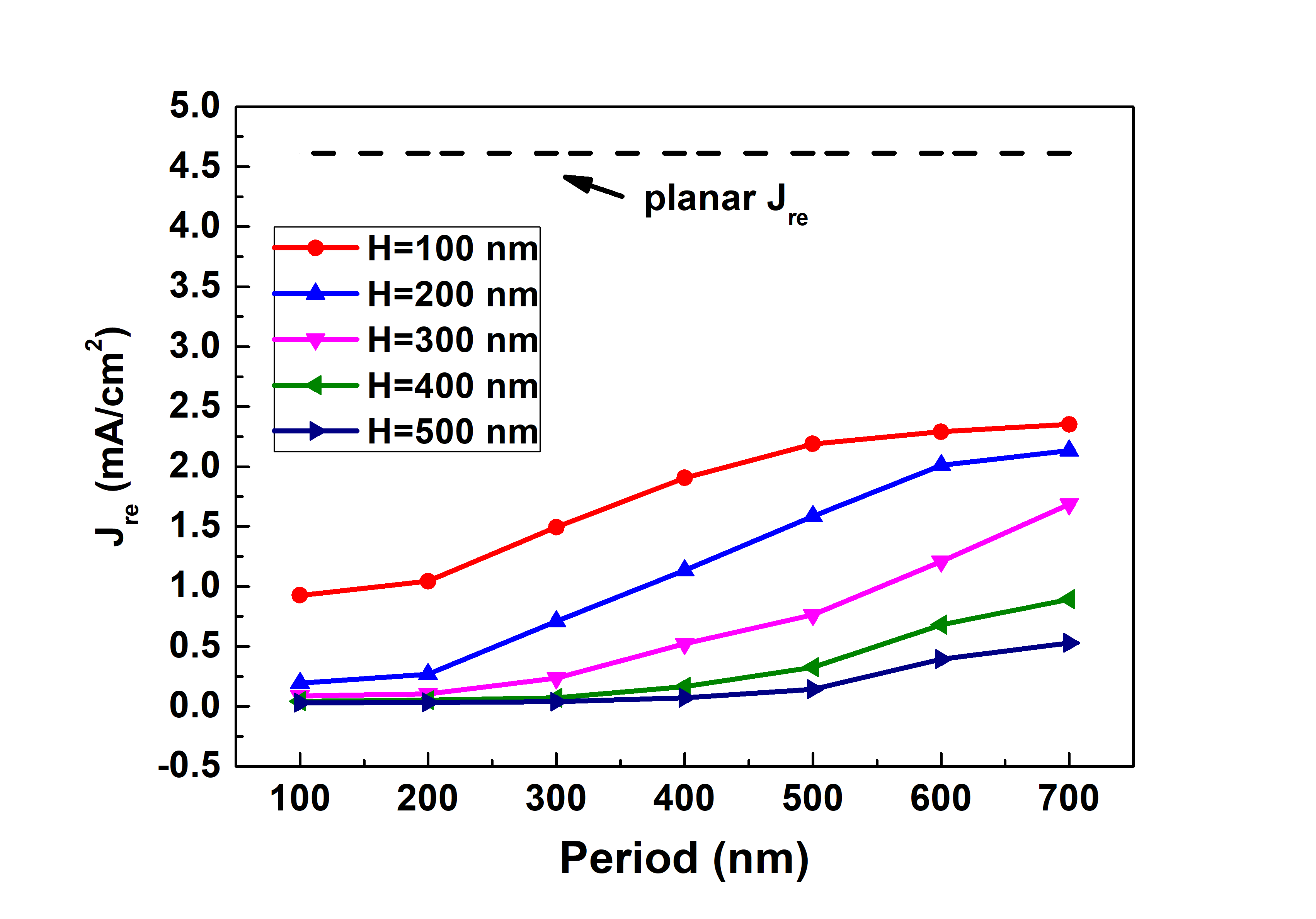}
\caption{Reflected photocurrent of TSCs with different texture structures.}
\label{fig:6}
\end{figure}

\begin{table}[!h]
\renewcommand{\arraystretch}{1}
\caption{Characteristic simulation and experimental parameters for perovskite/Si TSC.}
\label{table:1}
\centering
\begin{tabular}{c c c c c}
\hline
 & $J_{sc}$ [mA/cm$^2$] & $V_{oc}$ [V] & FF [$\%$] & PCE [$\%$] \\
\hline
Experiment\cite{Ref25} & 17.9 & 1.66 & 68 & 20.3 \\
Simulation & 17.9 & 1.66 & 71 & 21.3 \\ 
Experiment\cite{planar-Voc1.9} & 19.2 & 1.9 & 79 & 29.1 \\
Simulation (TJ opt.) & 17.9 & 1.91 & 75 & 25.8 \\
\hline
\end{tabular}
\end{table}

The bandgaps of perovskite and Si as the two main absorption layers are 1.68 eV and 1.12 eV, respectively. And the fitted value of $V_{oc}=1.9$~V is close to the experimental value of $V_{oc}$ shown in the current experiment\cite{planar-Voc1.9}. Ideally, for the bandgap of 1.68eV and 1.12eV for the two absorbing materials, the optimized $V_{oc}$ of each junction should be close to 1.33V and 0.762V ($\sim 0.35V$ smaller than $E_{g}/q$) or even higher. Hence, the theoretical $V_{oc}$ of the tandem cell should be close to 2.1V. To improve the PCE of the planar TSCs, both $J_{sc}$ or $V_{oc}$ still have room for improvement.Here we define $J_{\text{max}}$ as the theoretical maximum total photocurrent without any light reflection, and $J_{pho}$ as the actual photocurrent absorbed by the solar cell. Then the reflected photocurrent $J_{re}$ is defined as
\begin{equation}
J_{\text{re}} = J_{\text{max}} - J_{pho}.
\end{equation}
Ideally, in the two-junction tandem solar cell, if the current match is achieved, the short circuit current ($J_{sc}$) should be half of the absorbed photocurrent ($J_{pho}$), where $J_{sc}=J_{pho}/2$. In addition, the larger $J_{\text{re}}$ means more reflected light. 

For $J_{sc}$, by adding a triangular texture surface, the reflected photocurrent can be reduced effectively: compared with planar structures, TSCs with texture surfaces have a reflected photocurrent that is reduced from 4.5~mA/cm$^2$ to less than 2~mA/cm$^2$, as shown in Fig.~\ref{fig:6}. Additionally, different triangular structures reduce the reflected photocurrent, $J_{re}$, by varying amounts, necessitating the optimization of the texture structure. Before doing so, factors such as current matching of 2T TSCs, tunneling junction quality, and carrier nonradiative lifetime must also be considered. Ideally, reducing $J_{re}$ in the traditional single-junction solar cell is the key factor. However, in the tandem solar cell, if the increased $J_{pho}$ only occurs in either the perovskite layer or Si layer, resulting in unbalanced absorbed photocurrents on each junction, it will affect $J_{sc}$. Hence, if the currents are not matched, the case with a smaller reflected photocurrent may not lead to a higher $J_{sc}$ without the proper design layer thickness. Furthermore, even if the absorbed photocurrent is matched, if one absorbing layer's carrier lifetime is too short, causing its diffusion length to be shorter than the layer thickness, it will nonradiatively recombine before reaching the tunneling junction.

\begin{figure}[h]
\centering  \includegraphics[width=0.42\textwidth]{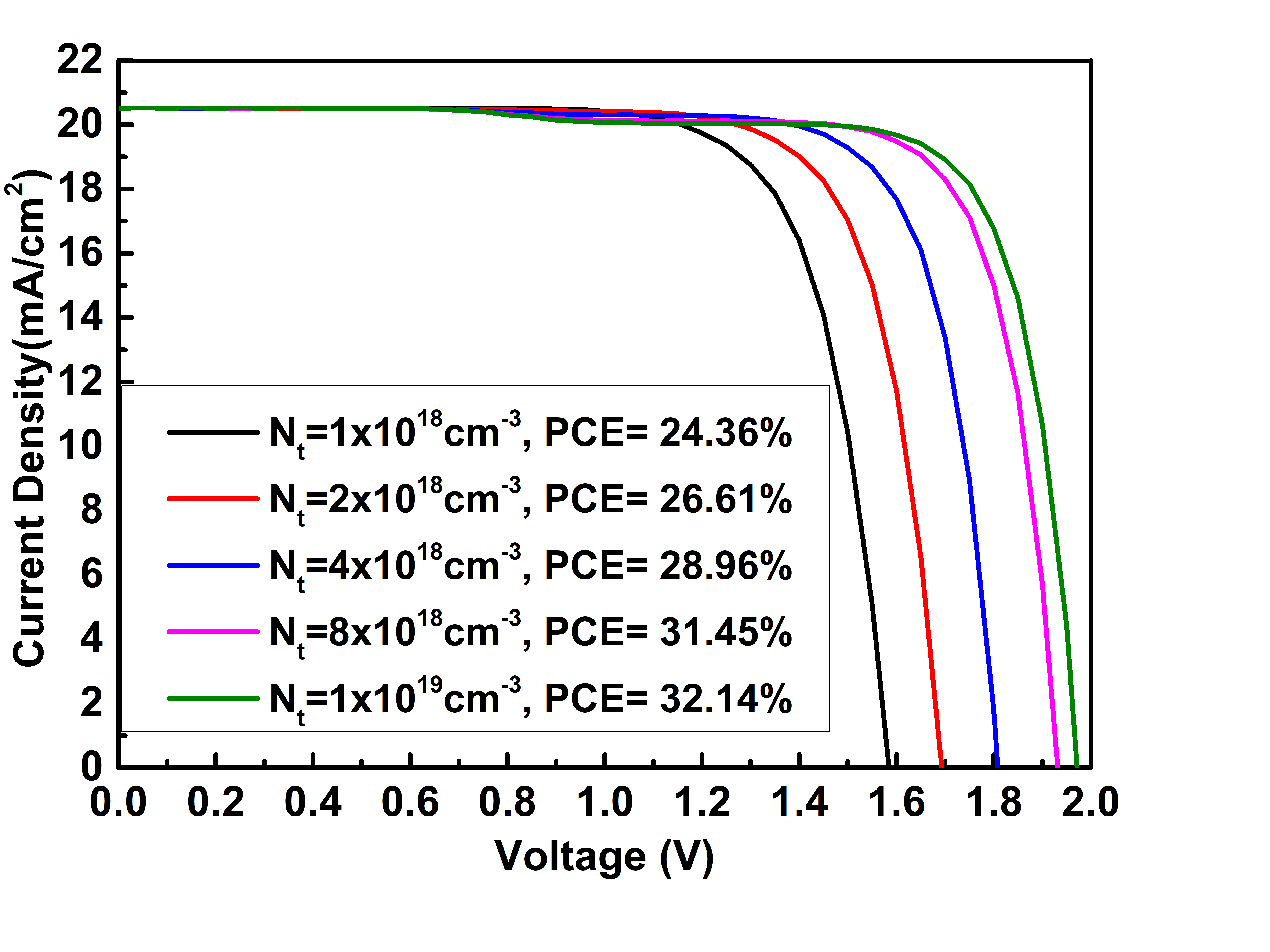}
\caption{$J$--$V$ curves for different values of defect-assisted tunneling concentration $N_t$.}
\label{fig:7}
\end{figure}

\begin{figure}[h]
\centering  \includegraphics[width=0.45\textwidth]{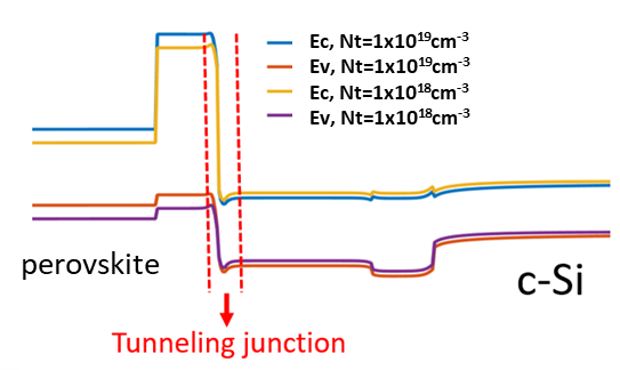}
\caption{Tunneling-junction band diagrams for $N_t = 10^{18}$~cm$^{-3}$ and $10^{19}$~cm$^{-3}$.}
\label{fig:8}
\end{figure}

Before studying the perovskite the layer thickness or its carrier lifetime, which affects $J_{sc}$ due to current match or nonradiative recombination, another important mechanism by which 2T TSCs affect the PCE is the tunneling junction between the two cells. Improving the tunneling junction by increasing the defect-assisted tunneling concentration $N_t$ can significantly improve $V_{oc}$, as shown in Fig.\ref{fig:7}. The setting of the parameter $N_t$ is established at the position of the tunneling junction, as shown in Fig.\ref{fig:8}. The tunneling junction is located at the interface between $\text{InO}_{\text{x}}$ and $\text{NiO}_{\text{x}}$.

\begin{figure}[h]
\includegraphics[width=0.45\textwidth]{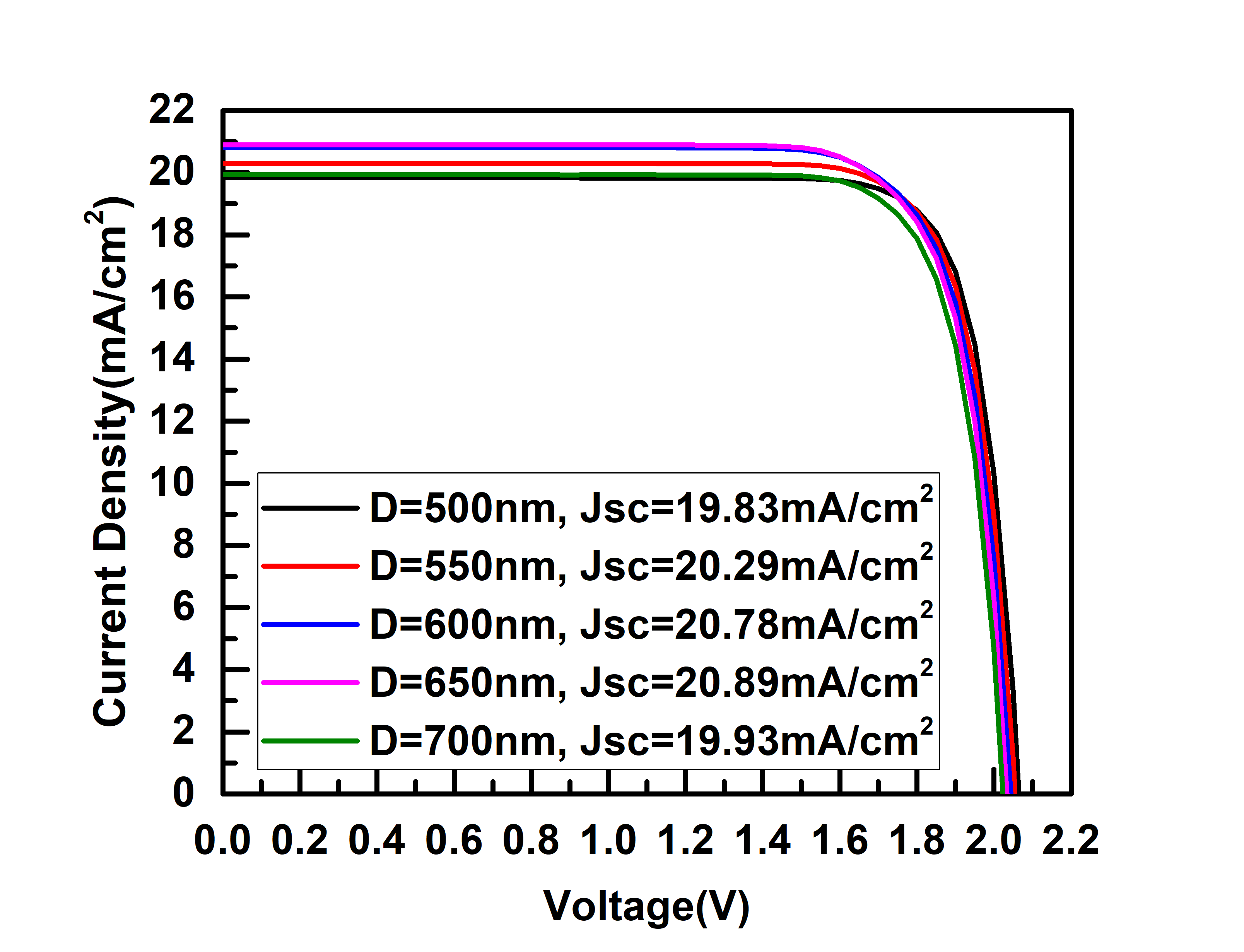}
\caption{ $J$--$V$ curves for perovskite thickness $D$ from 500~nm to 700~nm. $L=500$nm and $H=400$nm. }
\label{fig:9}
\end{figure}

\begin{figure}[h]
\includegraphics[width=0.40\textwidth]{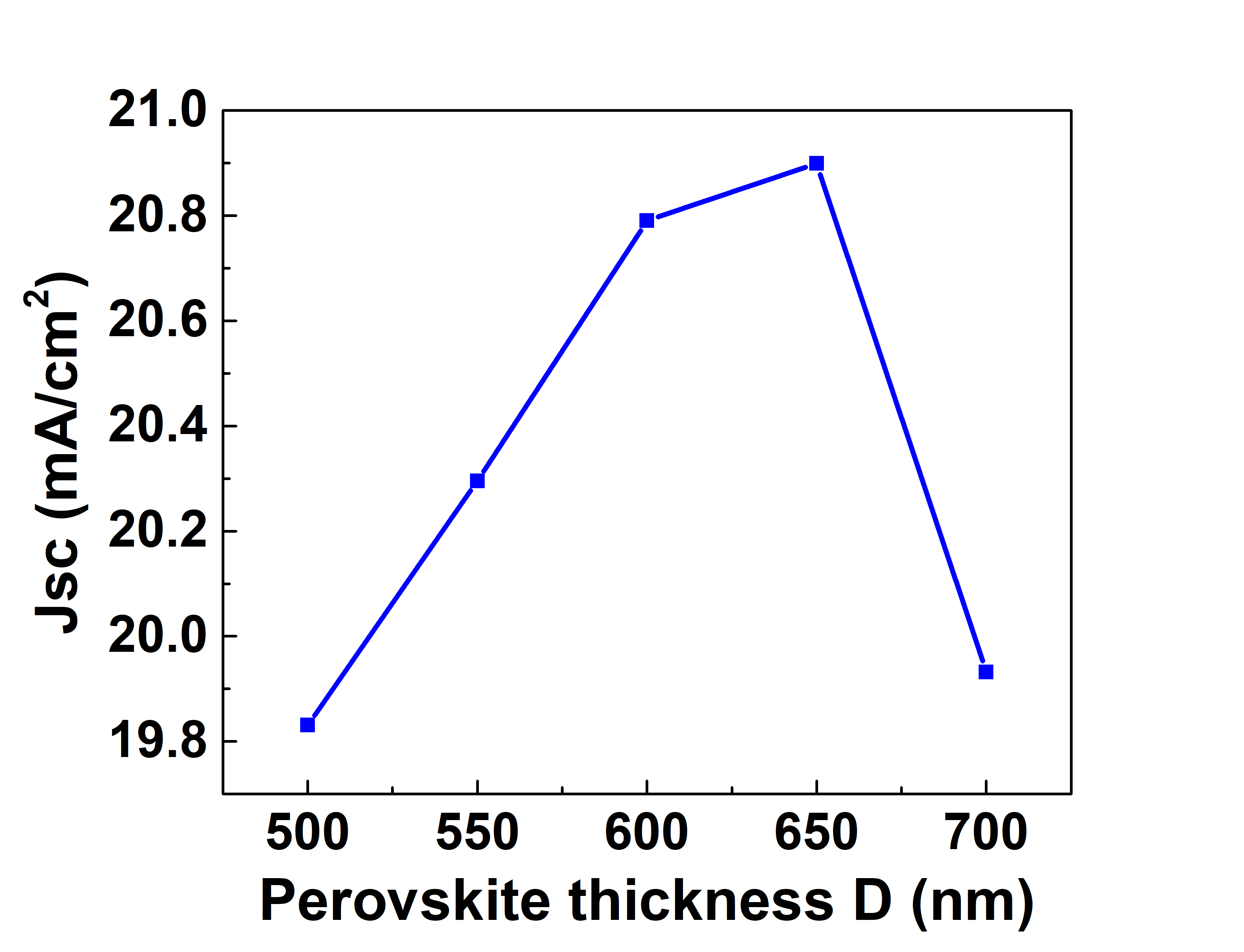}
\caption{Power conversion efficiency (PCE) for perovskite thickness $D$ from 500~nm to 700~nm. In these cases, $L=500$nm and $H=400$nm.}
\label{fig:10}
\end{figure}

 In this 2T textured perovskite/c-Si TSC, the c-Si thickness is fixed at 250~$\mu$m, while the perovskite thickness $D$ is adjusted from 500nm to 700nm. Figure~\ref{fig:9} shows the J-V curves for the texture case, where $L=500$ nm and $H=400$ nm. It illustrates that a perovskite layer that is either too thick or too thin results in a current mismatch, leading to a decrease in $J_{sc}$. It can be concluded that a perovskite thickness of 650nm is best for current matching; at this thickness, the device exhibits the highest $J_{sc}$ of 20.89 mA/cm$^2$, as shown in Fig.\ref{fig:10}. Hence, for different $L$ and $H$, we need to find the optimal perovskite layer thickness $D$ to match the current. The perovskite thickness for different periods $L$ and heights $H$ under current-matching conditions is provided in Table\ref{table:2}. Note that these $D$ values given are only under the photocurrent match condition. As mentioned earlier, the carrier lifetime in the perovskite layer and the effective shortest carrier escaping path (which is affected by the texture's shape) also play crucial roles in optimizing device performance.
  
\begin{figure}[h]
\centering  \includegraphics[width=0.49\textwidth]{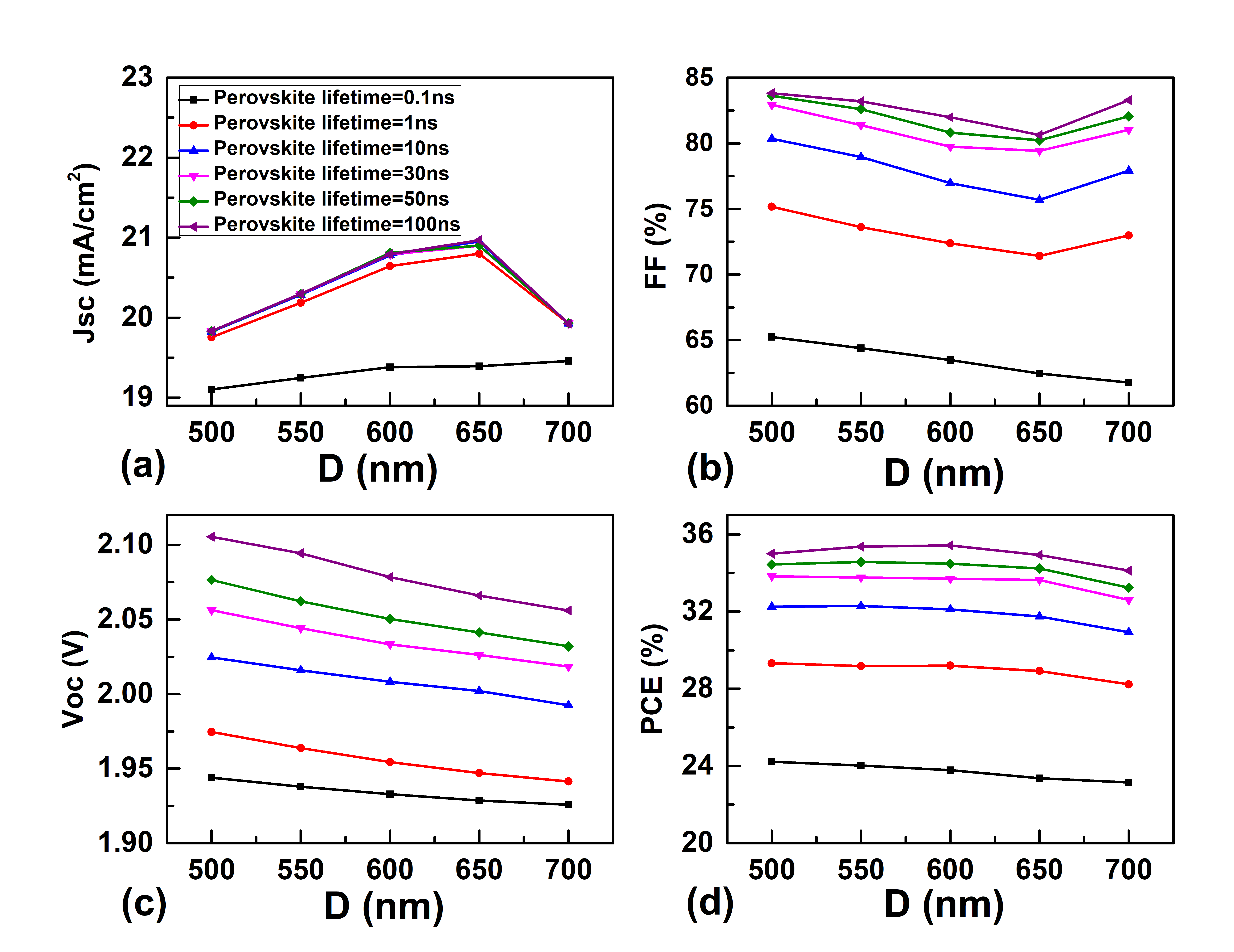}
\caption{Characteristic parameters for different values of perovskite thickness and carrier lifetime: (a) J$_{sc}$; (b) V$_{oc}$; (c) fill factor (FF); (d) PCE. In these cases, $L=500$ nm and $H=400$ nm.}
\label{fig:11}
\end{figure}

\begin{figure}[h]
\centering
  \includegraphics[width=0.4\textwidth]{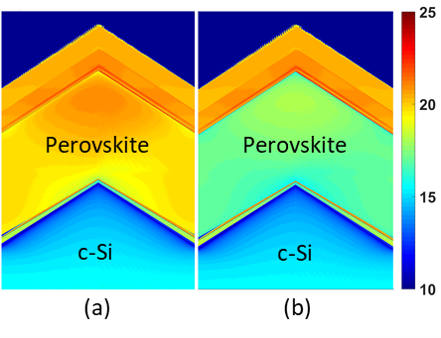}
\caption{Nonradiative recombination distribution (log scale) for perovskite carrier lifetimes of (a) 0.1~ns and (b) 100~ns. In these cases, $L=500$ nm and $H=400$ nm.}
\label{fig:12}
\end{figure}

To study the influence of carrier lifetime, the perovskite carrier lifetime is adjusted from 0.1ns to 100ns in structures with five different values of perovskite thickness mentioned above, to determine whether it can improve the current mismatch caused by inappropriate perovskite thickness. As shown in Fig.\ref{fig:11}(d) for the cases of $L=500$ nm and $H=400$ nm, except for the perovskite carrier lifetime of 0.1ns, longer carrier lifetimes all exhibit a similar trend in each characteristic parameter, and the optimum PCE is achieved with a perovskite thickness of 600nm. For the 0.1-ns case, the best PCE is obtained with a perovskite thickness of 500nm, which could be because a carrier lifetime of 0.1ns is too short compared to 100ns, resulting in significant nonradiative recombination within the perovskite layer, as shown in Fig.\ref{fig:12}(a). This also results in a significant decrease in $J_{sc}$, as shown in Fig.\ref{fig:11}(a).

\begin{table}[!h]
\renewcommand{\arraystretch}{1.5}
\caption{perovskite thickness of textured TSCs with different values of period $L$ and height $H$ under current-matching conditions (unit: nm).}
\label{table:2}
\centering
\begin{tabular}{c c c c c c c}
\hline
 & $L=200$ & $L=300$ & $L=400$ & $L=500$ & $L=600$ & $L=700$\\
\hline
$H=100$ & 650 & 675 & 700 & 700 & 700 & 675 \\
$H=200$ & 700 & 750 & 725 & 725 & 750 & 750 \\
$H=300$ & 750 & 800 & 775 & 675 & 625 & 700 \\
$H=400$ & 700 & 650 & 750 & 625 & 600 & 700 \\
$H=500$ & 700 & 675 & 550 & 625 & 600 & 700 \\
\hline
\end{tabular}
\end{table}

Having discussed the importance of current matching, we now seek to optimize the triangular texture structure with different value of height $H$ and period $L$. While optimizing the texture structure, we also adjust the perovskite thickness so that all structural devices are under current-matching conditions to ensure the best efficiency performance. The perovskite thickness of each device under current-matching conditions is given in Table~\ref{table:2} as we have mentioned earlier.

\begin{figure}[h]
\centering
\includegraphics[width=0.53\textwidth]{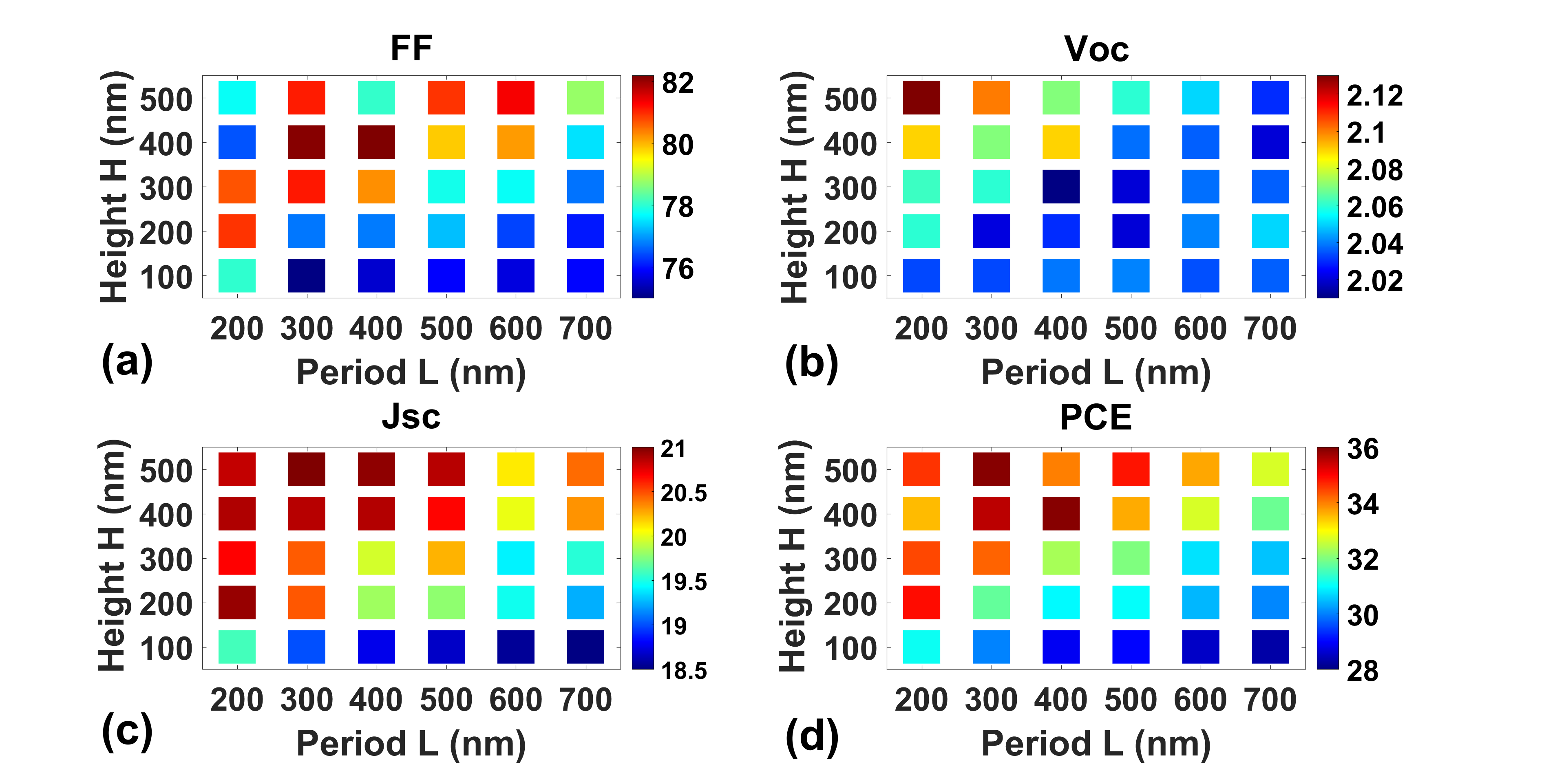}
\caption{Maps of characteristic parameters with different $H$ and $L$ under current-matching conditions: (a) FF; (b) $V_{oc}$; (c) $J_{sc}$; (d) PCE.}
\label{fig:13}
\end{figure}

As already mentioned, when the tip angle of the triangular texture structure is smaller, there will be less reflected photocurrent, and Fig.~\ref{fig:13}(c) shows that $J_{sc}$ in the electrical results also conforms to the same trend. The PCE map in Fig.~\ref{fig:13}(d) shows that the best efficiency is that of the structure with $H=400$~nm and $L=400$~nm.

\begin{figure}[h]
\centering
\includegraphics[width=0.49\textwidth]{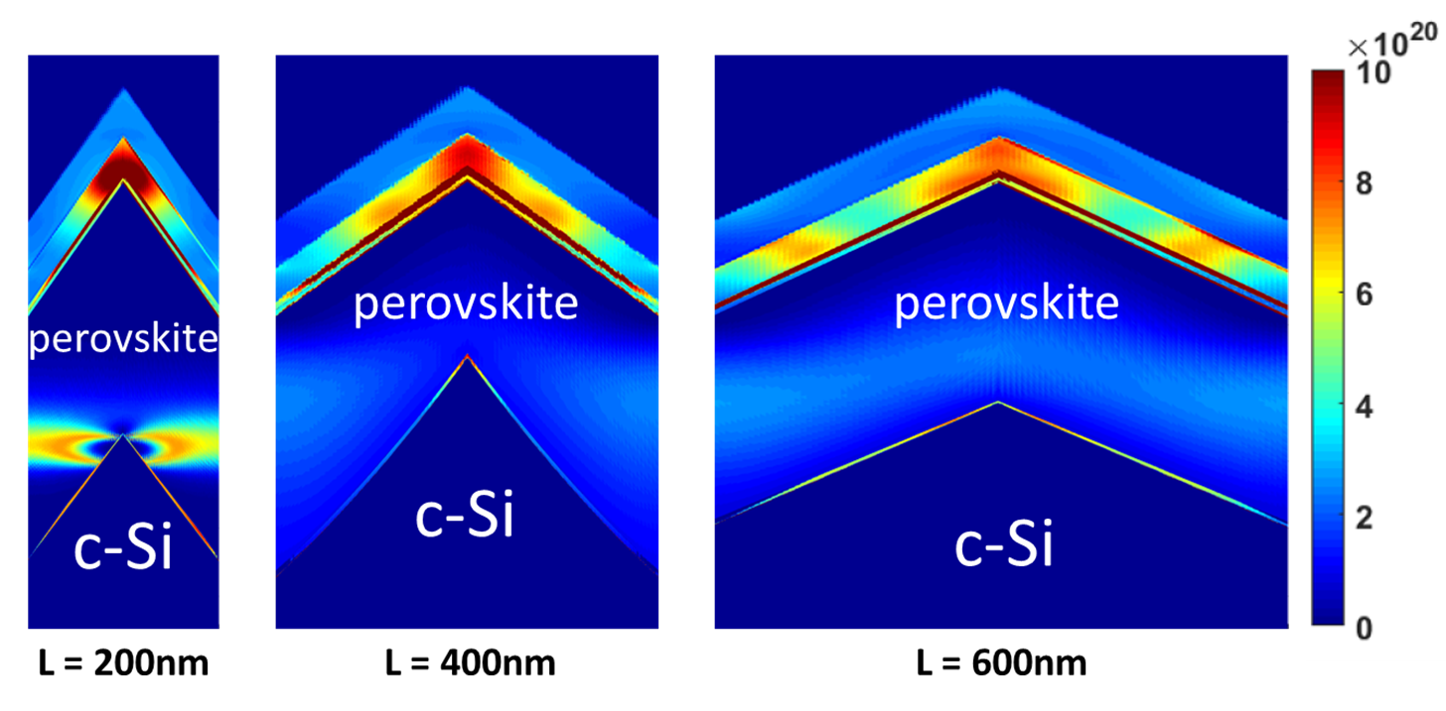}
\caption{Nonradiative recombination distribution under maximum power point for $H=400$~nm and different values of $L$ from 200~nm to 600~nm.}
\label{fig:14}
\end{figure}

Fixing $H=400$nm, we use three structures with $L=200$nm, 400nm, and 600nm to discuss how $L$ affects the PCE. Table~\ref{table:3} shows that the decrease in $J_{sc}$ for the structure with $L=600$~nm may be due to the larger reflected photocurrent mentioned earlier. There is no significant difference in $V_{oc}$ among the three structures. The main reason for the difference in PCE between the structures with $L=200$nm and 400nm is that the nonradiative recombination of the perovskite absorption layer at the maximum power point voltage for $L=200$nm is larger, as shown in Fig.\ref{fig:14}, which leads to a decrease in fill factor (FF), making the overall device efficiency lower than that of the structure with $L=400$~nm. Also, it can be seen that in the perovskite layer of the structure with $L=600$~nm, the overall nonradiative recombination is slightly larger than that for the structure with $L=400$~nm, causing the FF to decrease slightly by 2\%. Hence, although smaller $L$ gives the smallest reflection, it also results in larger nonradiative recombination in the perovskite layer near the corner region as shown in Fig.\ref{fig:14}. We found that the depletion field near the non-corner region at the maximum power point is very small, so nonradiative recombination is enhanced. This is also an important factor to consider.

\begin{table}[h]
\renewcommand{\arraystretch}{1.5}
\caption{Characteristic parameters for $H=400$~nm and period $L$ from 200~nm to 600~nm.}
\label{table:3}
\centering
\begin{tabular}{c c c c c}
\hline
 & $J_{sc}$ [mA/cm$^2$] & $V_{oc}$ [V] & FF [$\%$] & PCE [$\%$] \\
\hline
$L=200$~nm, $H=400$~nm & 20.88 & 2.09 & 76 & 33.5 \\
$L=400$~nm, $H=400$~nm & 20.87 & 2.09 & 82 & 35.9 \\ 
$L=600$~nm, $H=400$~nm & 20.01 & 2.03 & 80 & 32.6 \\
\hline
\end{tabular}
\end{table}

\begin{figure}[h]
\centering
\includegraphics[width=0.45\textwidth]{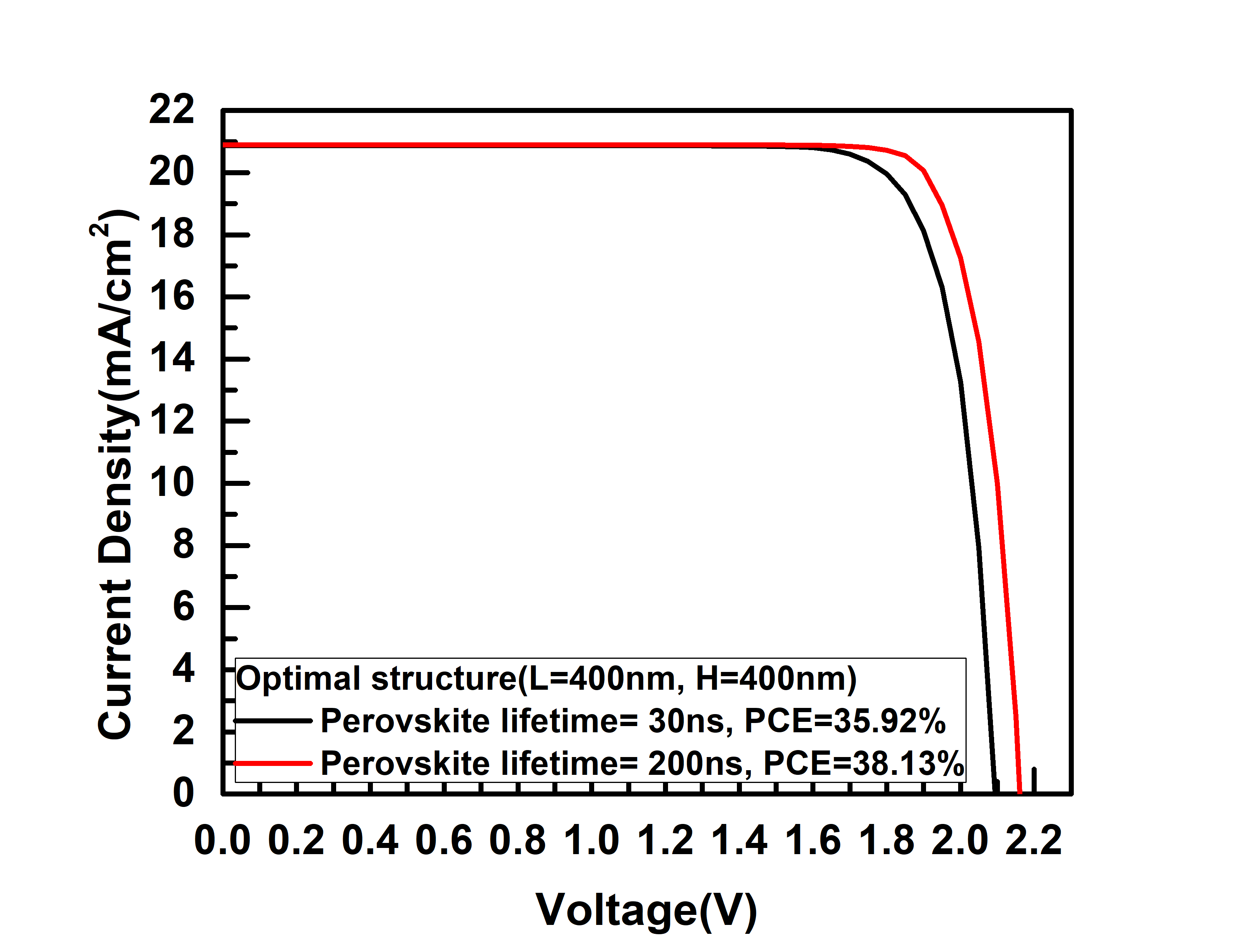}
\caption{$J$--$V$ curves of perovskite/Si TSCs with optimal structure for perovskite carrier lifetime of 30~ns and 200~ns.}
\label{fig:15}
\end{figure}

\section{Conclusion}

The main focus herein was exploring the optimization of the texture structure for 2T perovskite/Si TSCs. From the electrical results, the optimal structure with the highest PCE of 38.13\% is that with $H=400$~nm and $L=400$~nm. Furthermore, if the perovskite carrier lifetime can be increased from 30~ns to 200~ns, then $V_{oc}$ can be improved from 2.09~V to 2.16~V to obtain a higher PCE of 38.13\%, as shown in Fig.~\ref{fig:15}. Therefore, achieving the anticipated efficiency levels outlined in this research depends on upcoming efforts to construct the suggested texture structure. It is essential to focus on improving the quality of perovskite materials and enhancing the tunneling junction. In the future, 2T perovskite/Si TSCs could reach an efficiency of $\sim$38\%.

\section{DATA AVAILABILITY}
The data that support the findings of this study are available
from the corresponding author upon reasonable request. The simulation tool with all functions used in this software can be download in https://yrwu-kw.ee.ntu.edu.tw for free in academic use.

\section*{Acknowledgment}
This work was supported by the National Science and Technology Council (Grant Nos.\ 112-2221-E-002-214-MY3 and 112-2221-E-002 -215-MY3, and 112-2923-E-002-002).

\nocite{*}
\bibliography{IEEE}

\end{document}